\newcommand{\beq}{\begin{equation}}
\newcommand{\eeq}{\end{equation}}
\newcommand{\beqa}{\begin{eqnarray}}
\newcommand{\eeqa}{\end{eqnarray}}

\newcommand{\br}{{\bf r}}

\renewcommand{\lambda}{\ell}

\newcommand{\0}{\Delta_0}
\newcommand{\1}{\Delta_1}

\documentstyle[aps,prl,multicol,fancyheadings,epsfig,epsf,amsbsy,amssymb,amstex]
{revtex}
\begin{document}
\title{Collective Mode in a $d_{x^2-y^2} + id_{xy}$ Superconductor }
\author{  A. V. Balatsky$^{1}$,  P. Kumar$^2$ and J. R.
Schrieffer$^{3}$}
\address{
$^1$Theoretical Division, 
Los Alamos National Laboratory, Los Alamos, NM 87545, USA.\\
$^2$ Department of Physics, PO Box 118440, University of Florida,
Gainesville FL 
32611   \\
$^3$ NHMFL, Florida State University, 1800 E. Paul Dirac Dr.
Tallahassee, FL 
32310\\
}
\date{Received \today }

\maketitle

\begin{abstract}
We consider a superconducting state with a mixed symmetry order
parameter components, e.g. $d+is$ or $d+id'$ with $d'= d_{xy}$. We
argue for the existence of the new orbital magnetization mode which
corresponds to the oscillations of relative phase $\phi$  between two
components around an equilibrium value of $\phi = \frac{\pi}{2}$. It
is similar to the so called ``clapping'' mode in superfluid
$^3He-A$. We estimate the frequency of this mode $\omega_0(B,T)$
depending on the field and temperature for the specific case of
magnetic field induced $d'= d_{xy}$ state.  We find that this mode is
{\em tunable} with an applied magnetic field with $\omega_0(B,T)
\propto B \0$, where $\0$ is the magnitude of the d-wave order
parameter. We also estimate the velocity $s(B,T)$ of this mode.
Pacs Numbers: 74.20.-z, 74.25.Nf
\end{abstract}

\vspace*{1.0cm}
\begin{multicols}{2}

\columnseprule 0pt

\narrowtext

The order parameter in high-Tc superconductors is widely believed
\cite{Annett1,wallman93,hardy93,shen93,Tsuei98} to be of a $d_{x^2 -
y^2}$ symmetry. However more careful consideration indicates that the
symmetry of the state in high-Tc might be lower in a number of
cases. The symmetry allows the secondary components to be generated
whenever there is a perturbing field.  These secondary components
($is$ or $id_{xy}$) generation has been addressed in recent literature
for the case of inhomogeneity due to wall scattering
\cite{Matsumoto1,Sauls1,Green1,Kashiwaya1,Tanuma1,Kuboki1} or due to
vortex texture \cite{Rama1,Himeda1,Soininen1,Kim1,Ichioka1}.
Similarly, the generated $id'$ component near magnetic impurity and
spontaneous violation of time reversal symmetry with global $d+id'$
has also been discussed
\cite{Balatsky1,Graf1,Grimaldi1}. Another possibility for a global
$d+id'$ phase 
has been pointed out\cite{KT,Laugh1,Balatsky2} where the external
magnetic field 
applied to two-dimensional $d$-wave superconductor generates $id'$. 

 Direct experimental determination of the second superconducting
component nevertheless remains a substantial challenge, see
e.g. \cite{Green1}.  The smallness of induced component and the large
superconducting response of the underlying $d $ component makes the
detection of the secondary component difficult.

We present here an excitation which constitutes a direct test of the
induced component of the order parameter. This excitation is uniquely
tied to the existence of the secondary {\em out of phase} component of
the order parameter. Consider the most general situation of a {\em
complex} order parameter which can be generally written as $\0 + i\1$
where $\0$ is the original $d_{x^2-y^2}$ component and $\1$ is the
induced $s$ or $d_{xy}$ component which is orthogonal to the initial
$d_{x^2-y^2}$ state. Define the global (Josephson) phase of the order
parameter $\nu$ and the relative phase $\phi$ as:
\beqa
\Delta = [|\0| + exp(i\phi(\br))|\1|]exp(i\nu(\br))
\label{def}
\eeqa
 If the order parameter $\Delta$ is to be defined at the Fermi
surface, then the functions $\Delta_i, i=0,1$ implicitly contain the
angular dependence.  The global phase $\nu(\bf{r})$ can be position
dependent and even singular as is the case for the vortex
configuration. We will focus on the relative phase $\phi$. Its
dynamics by definition is related to the appearance of the secondary
component $\1$. While the equilibrium value of the relative phase in
most cases is argued to be $\pm \frac{\pi}{2}$ resulting in a
violation of the time reversal symmetry, $\phi$ may oscillate around
this value and represent a new collective mode that can be used to
detect the induced component. Similar relative phase mode has been
considered before in context of two band superconductors
\cite{Leggett2} and heavy fermions \cite{Kumar1} and in \cite{Brusov1}, where   
the collective modes of d-wave superconductor in zero field were considered. 

This collective mode is akin to a spin wave in the orbital moment of
the superconducting ground state. The dispersion of this mode has a
gap similar to the Larmour frequency in case of spins.  As has been
shown in
\cite{Kumar1}, the 
relative phase oscillations also can be viewed as an internal
Josephson effect, similar to the clapping mode in $^3He-A$
\cite{Leggett1} or to the relative phase dynamics in a 2-band
supercondutor\cite{Leggett2}.  In the sense of a magnetic excitation,
that can respond to a time dependent magnetic field in a resonant
manner, this mode is also comparable to the longitudinal NMR in
$^3He-A$ \cite{Leggett1}.  The details of the dynamics are clearly
model dependent.

To be specific we will focus on the field induced $\1 = id_{xy}$
secondary component in the bulk. In this case quasiparticle spectrum is fully 
gapped in the field
with the minimal gap vanishing at zero field. The $d+id'$ state breakes time
reversal symmetry and has a finite magnetic moment $M_z$ perpendicular
to the superconducting plane\cite{Laugh1,Balatsky2}. The mode
discussed above in this case will correspond to the {\em longitudinal}
oscillations of the condensate magnetic moment around its equilibrium
value. The frequency of this mode will contain information about the
details of the coupling between components and will provide insight
into the overall order parameter structure. Below we will focus  on a 2d 
superconductor at
the fields $H \simeq H_{c2}(T)$ and therefore we will ignore small
($O(H-H_{c2})$) difference between induction $\bf{B}$ and applied
field $\bf{H} ||z$.

We find the resonance frequency for the {\em relative} phase
oscillations of $\phi$. It turns out to be a gapped propagating mode
with:
\beqa
\omega^2(B,k) = \omega^2_0(B,T) + s^2_{ij}(B,T)k_ik_j, \nonumber\\ 
\omega_0(B) \simeq \frac{\eta B}{N(0)} \0(B,T)
\label{mode1}
\eeqa
\beqa
 s^2_{ij} = \delta_{ij} s^2 \ \
\label{vel1}
\eeqa
Here $s = (a + b B^2)\Delta^2_0$ is the mode velocity, $a,b$ are some constants and  $\eta$ is a constant, a measure of
the strength of the interaction, which we discuss in more detail
below; $N(0)$ is the Density of States (DOS) at the Fermi level; both
$i,j$ refer to in-plane coordinates $x,y$. This mode is a longitudinal
magnetization oscillation $\delta M_z(t,r)\propto \delta M_z exp(i
\omega(B,k)t 
-ikr)$ and is {\em tunable by the external field}. There are, in
effect, two consequences arising from a non-zero $\eta$ which may be
used to estimate its magnitude.  In the presence of a secondary order
parameter $\Delta_1 = id_{xy}$, the temperature dependent upper
critical field $H_{c2}(T)$ has an additional contribution which is
quadratic in $(1-T/T_c)$.  This results in an upward curvature with a
scaling field which depends on $\eta$.  Thus the upward curvature in
$H_{c2}(T)$ and the resonance frequency should be related in that they
are both derivable form the same energy scale.

\begin{figure}
\epsfysize=2.0in
\centerline{{\epsfbox{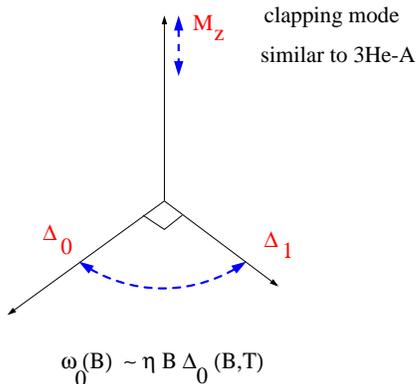}}}
\caption[]{Schematic representation of the longitudinal magnetization 
oscillations caused by the oscillation of the relative phase angle
between $\0$ and $\1$ with an equilibrium value $\pi/2$. The frequency
of this mode is linear in the B field and is also linear in the
magnitude of the $\0$. This mode is therefore tunable by the field and
by temperature. We estimate $\omega_0(B,T)
\simeq \1(B,T)$ and is in the Kelvin range.}
\label{fig_clap}
\end{figure}

We now will prove the existence of the ``clapping mode'' for $d+id'$
state. The free energy has the standard form :
\beqa
F = F_0 +F_1 + \frac{B^2}{8\pi}
\label{F}
\eeqa
\beqa
F_0 = \frac{a_0}{2}|\0|^2 + \frac{b_0}{4}|\0|^4 +
\frac{K_{ij}}{2}|D_i\0D_j\0^*|
\label{F0}
\eeqa
\beqa
F_1 = \frac{a_1}{2}|\1|^2 +\frac{K'_{ij}}{2}|D_i\1D_j\1^*|
\label{F1}
\eeqa
\beqa
F_{int} = \frac{\eta}{2}[D_x,D_y]\0\1^*
\label{Fint}
\eeqa
where $\0,\1$ are $d$ and $d'$ components of the order parameter, $a_0
= 
\alpha_0(T/T_0 - 1)$ , $T_0$ is the ordering temperature for $\0$. The 
corresponding $a_1>0 = N_0$ is always positive, as are
 $b_{0}>0$ and $K_{ij},K'_{ij}>0$.  $D_i =
\partial_i - i\frac{2e}{c}A_i$, with ${\bf B} = {\bf \nabla} \times
{\bf A}$.  For simplicity we take $K_{ij}= K \delta_{ij}$ and
similarly for $K'$ term.  Because we are near $H_{c2}$, the flux
expulsion is nearly non-existent and $B$ and $H$ are nearly equal. The
interaction term in this form has been proposed earlier
\cite{KT,Balatsky2}. One can see that the GL free energy is indeed a
scalar. The interaction term, using $[D_x,D_y] = i e B$ can be written
as
\beqa
F_{int} = i eB\frac{\eta}{2} \0\1^* + h.c.
\label{Fint2}
\eeqa
which corresponds to a coupling of the magnetic field with an {\em 
intrinsic} orbital moment along z-axis:$<M_z> = \frac{i\eta }{2}\0\1^*
+ h.c.$.

Up to now the work on secondary component has been focused on the
equilibrium solution for a particular model for induced component, say
in $d+is$ or $d+id'$ state.  Here we will assume that the amplitude
for the secondary $\1$ has been developed and we will look at the
relative phase $\phi$ oscillations only.  It is convenient to
introduce the respective phases of each component:
\beqa
\0 = |\0| exp(i\phi_0), \ \  \1 = |\1| exp(i\phi_1)
\label{phases}
\eeqa
which are related to the introduced above $\nu = \phi_0, \phi = \phi_1
-
\phi_0$. Derivation of the Eq.(\ref{mode1}) then proceeds as follows.
We use the Josephson relations for each component:
\beqa
\dot{\phi_i} = \frac{\partial F}{\partial N_i} = \mu_i,  \ \  \dot{N_i}
= 
-\frac{\partial F}{\partial \phi_i}, i = 0,1
\label{Josephson}
\eeqa
Where $\mu_i$ are chemical potentials for particles in $\0,\1$
condensate and similarly, $N_i$ are conjugated number of particles. In
the equilibrium, where $\mu_0 = \mu_1 = \mu$, we find for a relative
phase motion:
\beqa
\ddot{\phi} = - [\frac{\partial \mu_0}{\partial N_0}+\frac{\partial 
\mu_1}{\partial N_1}-2\frac{\partial \mu_0}{\partial N_1}]
\frac{\partial 
F}{\partial \phi}
\label{Josephson2}
\eeqa
where in the above we have taken into account the fact that although
$\mu_0 =
\mu_1$ 
their derivatives are different.  The term in the brackets is on the
order of $N(0)$: $[\frac{\partial
\mu_0}{\partial N_0}+\frac{\partial 
\mu_1}{\partial N_1}-2\frac{\partial \mu_0}{\partial N_1}] = \rho^{-1}
\simeq 
N(0)$.  For the general values of the relative phase $\phi$ the terms
in free energy $F$ that contribute to the derivative in
Eq.(\ref{Josephson2}) are $F_{int}$, Eq.(\ref{Fint}) and terms with
gradients $K_{ij},K'_{ij}$.
\beqa
F_{int} = \eta B_z |\0||\1| \sin \phi
\label{phi}
\eeqa
 with minimum $F$ reached at $\phi = -\pi/2 sgn (B_z)$.  In general,
there is an additional term proportional to the squares of the two
(the bare and the induced) order parameters.  However it contributes
little new to the physics of the problem. Its effects are largely
quantitative.  We stress here that the phase $\phi_0$ of $\0$ is not
constant in the external field in the mixed state. Nevertheless the
$F_{int}$ in Eq.(\ref{phi}) depends on the relative phase only.

\begin{figure}
\epsfysize=2.0in
\centerline{{\epsfbox{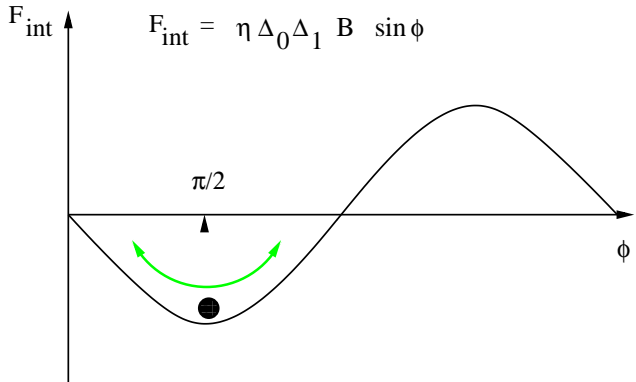}}}
\caption[]{The $F_{int}$ profile as a function of the relative phase
angle $\phi$ between $\0$ and $\1$ is shown. Although the minimum is
reached at $\phi = \frac{\pi}{2}$ the finite stiffness for $\phi$
oscillations leads to the finite frequency mode at $\omega_0(B,T)$. We
have ignored the dependence on the sign of $B$ in the figure.  }
\label{fig_clap2}
\end{figure}

We find:
\beqa
\ddot{\phi} =  \frac{1}{\rho}\eta |\0||\1| B_z \cos \phi - s^2\nabla^2
\phi
\label{Josephson3}
\eeqa

Here $s$ is given by Eq.(\ref{vel1}).   Minimizing the free energy
Eq.(\ref{F},\ref{F0},\ref{F1}) with respect to the magnitude from this
equation, Eq.(\ref{mode1}) follows immediately. The velocity $s$ is field dependent and also can be used in
experiments to detect the presence of $id'$ component.

We can estimate the magnitude of the energy gap up to a $O(1)$
prefactors:
\beqa
\omega_0(B,T) \simeq |\Delta_1(B,T)| \simeq  \frac{\eta 
B}{N(0)} \0(B,T)
\label{estimate1}
\eeqa
as one can easily see from minimization of the total $F$ with respect
to $\1$.  For estimate for the coupling constant $\eta$ and magnitude
of $\1$ see
\cite{Balatsky2}. We estimate thus  
\beqa
\frac{\omega_0(B,T)}{\0(B,T)}  \simeq
(\frac{a^2}{\xi^2})(\frac{H}{H_{c2}}) 
\label{estimate2}
\eeqa
This result turns out to be very similar to the ``clapping mode'' in
$^3He-A$, where frequency was found to scale with the gap in the whole
temperature range as well \cite{3He}.  We also find asymptotics in
$H$:
\beqa
\omega_0 \sim \left\{
\begin{array}{cc}
 H/H_{c2}  &  H \leq H_{c2}\\
 \sqrt{H_{c2}-H} &  H\rightarrow H_{c2}\\
\end{array}
\right.
\label{estimate3}
\eeqa
and we note immediately that the mode frequency $\omega_0 \ll \0$
Therefore this mode will be sharp. The damping coming from the
low-lying quasiparticles in the nodes of d-wave gap will not affect
this mode because the phase space for the decay will be small.   

In above estimates we assumed that the field induced component $id'$
will be present even outside the GL region for which the above
formulas have been derived. We assumed that once the field induced
component is present in GL region it will persist to a lower
temperatures. Realizing the limitations of estimates made beyond GL
regime we present Eq.(\ref{estimate3}) as an order of magnitude
estimate only. From Eq.(\ref{estimate2}) it follows that the gap
$\omega_0$ can be made in the range of $0.1-1 K (2-20 GHz)$ in the
field $H = 1-10 T$ similar to the $\1$ estimates
\cite{Balatsky2}. 

The longitudinal oscillations of magnetization $M_z$ will also lead to
the resonance at $\omega_0(B,T)$ in the AC susceptibility of the
superconducting state. Experimentally the proposed clapping mode can
be observed in NMR in the field. Another possible method of detecting
this resonance is the ultrasound attenuation (taken at lower fields),
used previously in heavy fermions to investigate the excitations of
the order parameter. For any of the resonance techniques used to
search for the ``clapping mode'' resonance at $\omega_0(B,T)$
Eq.(\ref{estimate2}) the vortex cores contribution would be the
biggest source of background. We note however that $\omega_0(B,T)$ can
still be measured if the resonance is sharp enough.

In conclusion, we considered the relative phase oscillation mode that
can be used to detect the secondary component in the {\em time
reversal violating } superconducting state, such as $d+id'$ or $d+is$.
Two components of the order parameter are characterized by respective
amplitudes and phases. The relative phase between two components can
oscillate around its equilibrium value $\phi =
\pm \frac{\pi}{2}$. This mode is very similar to the ``clapping mode''
in superfluid $^3He-A$. For a specific model we choose the case of
field induced $id'$ component. We show how the relative phase mode
frequency is governed by external magnetic field and d-wave gap
magnitude $\omega_0(B,T) \propto B \0$ and is therefore {\em tunable}
by external magnetic field. The scale for $\omega_0$ and mode velocity
$s$ is set by the magnitude of the induced gap $\1$ and we estimated
the frequency for different fields, Eq.(\ref{estimate2}). Among other
probes this mode could be experimentally detected by investigating the
ac magnetic susceptibility and possibly by ultrasound attenuation in
the in the mixed state as a function of applied field $H$.


 Work at Los Alamos was sponsored 
by the US DOE.

\end{multicols}

\end{document}